\documentclass[%
 reprint,
 amsmath,amssymb,
 aps,
 prl,
]{revtex4-2}
\usepackage{siunitx}
\usepackage[export]{adjustbox}
\usepackage{graphicx}% Include figure files
\usepackage{dcolumn}% Align table columns on decimal point
\usepackage{bm}% bold math
\usepackage{subcaption}
\usepackage{caption}
\DeclareSIUnit\gauss{G}
\DeclareSIUnit\ppm{ppm}
\DeclareSIUnit\phonon{phonon}
\usepackage{comment}
\usepackage{xr}
\usepackage{hyperref}
\begin{document}

%\preprint{APS/123-QED}
%TC:ignore

\title{Heating of a trapped ion induced by dielectric materials}

\author{Markus Teller$^{1}$}

\author{Dario A. Fioretto$^{1}$}
\author{Philip C. Holz$^{1,2}$}
\author{Philipp Schindler$^{1}$}
\author{Viktor Messerer$^{1}$}
\author{Klemens Sch\"uppert$^{1}$}
\author{Yueyang Zou$^{1}$}
\author{Rainer Blatt$^{1,3}$}
\author{John Chiaverini$^{4,5}$}

\author{Jeremy Sage$^{4,5}$}
\author{Tracy E. Northup$^{1}$}
\affiliation{$^{1}$Institut für Experimentalphysik, Universität Innsbruck, Technikerstrasse 25, 6020 Innsbruck, Austria}
\affiliation{$^{2}$Alpine Quantum Technologies GmbH, Technikerstrasse 17/1, 6020 Innsbruck, Austria}
\affiliation{$^{3}$Institut für Quantenoptik und Quanteninformation, Österreichische Akademie der Wissenschaften, Technikerstrasse 21 A, 6020 Innsbruck, Austria}
\affiliation{$^{4}$Lincoln Laboratory, Massachusetts Institute of Technology, Lexington, MA, USA}
\affiliation{$^{5}$Massachusetts Institute of Technology, Cambridge, MA, USA}

\date{\today}% It is always \today, today,

\begin{abstract}
Electric-field noise due to surfaces disturbs the motion of nearby trapped ions, compromising the fidelity of gate operations that are the basis for quantum computing algorithms. We present a method that predicts the effect of dielectric materials on the ion's motion. Such dielectrics are integral components of ion traps. Quantitative agreement is found between a model with no free parameters and measurements of a trapped ion in proximity to dielectric mirrors. We expect that this approach can be used to optimize the design of ion-trap-based quantum computers and network nodes.
\end{abstract}
\maketitle
%TC:endignore
%\section{Introduction}

In a trapped-ion quantum computer, entangling gate operations rely on a shared motional degree of freedom~\cite{CiracZoller1995,PhysRevLett.82.1835}.
For high-fidelity entanglement, as is now standard in prototype quantum computing experiments~\cite{Schmidt-Kaler2003,PhysRevLett.94.153602,Erhard2019}, the ions' motional state must be preserved during a gate operation~\cite{PhysRevLett.82.1835,Monroe95a}.
This requirement may be compromised by electric-field fluctuations, which in a trapped-ion experiment couple to the motion of the ions and increase their temperature~\cite{Monroe95a,Turchette2000}.
%These fluctuations originate from nearby surfaces and therefore are often referred to as surface noise~\cite{Daniilidis2014,RevModPhys.87.1419}.
Electric-field noise presents a challenge in particular for scaling up the number of quantum bits while maintaining high-fidelity gate operations~\cite{RevModPhys.87.1419,brown2020materials}.

While electric-field noise is a topic of much current research, we lack a complete understanding of the physical mechanisms behind it~\cite{RevModPhys.87.1419}.
In recent years, Johnson noise, patch potentials and surface adsorbates have been investigated as noise sources~\cite{Daniilidis2014,PhysRevA.99.063427,PhysRevA.84.053425,PhysRevA.87.023421}. It has been pointed out that dielectric materials may be a significant noise source even for thin layers, such as in the case of surface oxides and adsorbates~\cite{Kumph_2016,Rubenstein}.
%, and a recent experimental study provides support for the hypothesis of surface noise due to dielectric oxide layers on top of a metallic electrode~\cite{}.

Here we present and test a method that predicts ion motional heating due to the proximity to a dielectric. We take advantage of an experimental setup with integrated optics whose distance to a trapped ion can be varied, in which the dielectric makes the dominant contribution to electric-field noise.  Our measurements show quantitative agreement with our predictive model, which has no free parameters. %Due to the ubiquity of dielectrics in ion traps, we anticipate that our approach will enable the ion-trap community to make a broad assessment of the contribution of dielectrics to ion heating in existing traps.
Since dielectrics are ubiquitous in ion traps as integrated optics~\cite{Guthoehrlein2001,Eschner01,PhysRevLett.89.103001,VanDevender,Kim2,Steiner2013,Niffenegger2020,Mehta2020}, as trap substrates~\cite{SurfaceTrapReview,Cho2015}, and as surface contamination, we anticipate that our approach will enable a broad assessment of the contribution of dielectrics to ion heating in existing traps. Moreover, it will also allow researchers to minimize the motional heating of future traps during the design and fabrication phases, an important step for the development of quantum computers and for quantum network nodes, which require efficient optical interfaces.

First, we summarize the formalism for predicting the electric-field noise at the position of an ion due to a dielectric with arbitrary geometry.
\begin{figure}
\centering
\includegraphics[scale=0.75]{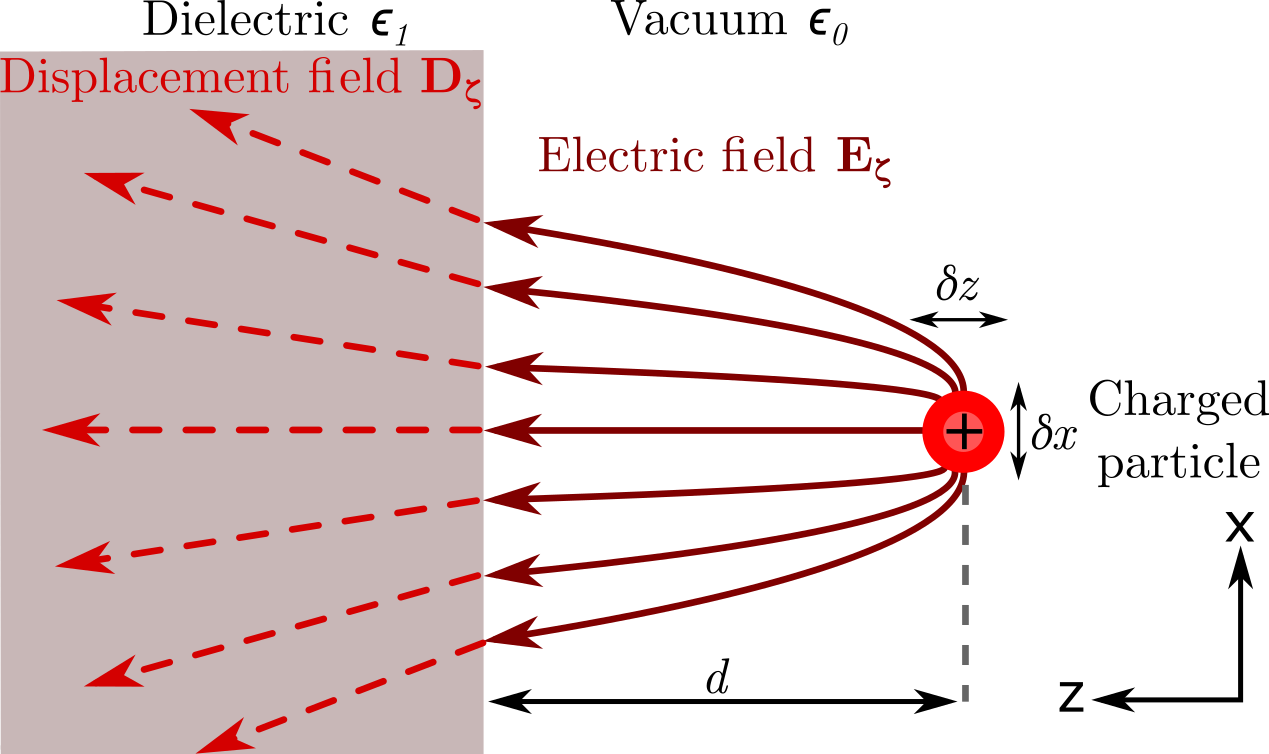}
\caption{%Schematic representation of the key elements of the formalism. 
	A charged particle is placed a distance $d$ from a dielectric material with permittivity $\epsilon_1$. The particle oscillates in vacuum (permittivity $\epsilon_0$) with amplitude $\delta \zeta\ll d$, with $\delta \zeta \in\{\delta x,\delta z\}$. The charge generates an electric field $\mathbf{E}_\zeta$ outside the dielectric and a displacement field $\mathbf{D}_\zeta$ inside.}\label{fig:Schematic}
\end{figure}
The formalism is based on the derivation of Ref.~\cite{Kumph_2016}, which considers the case of infinite planes, and which we have adapted for finite volumes of arbitrary shape; a sketch of the key elements is depicted in Fig.~\ref{fig:Schematic}.
We consider a charged particle with mass $m$ that oscillates with frequency $\omega$  in vacuum (permittivity $\epsilon_0$) near a dielectric material.
We assume that the distance $d$ between the dielectric and the particle is significantly larger than both the oscillation amplitude parallel to the surface, $\delta x$, and the oscillation amplitude perpendicular to the surface, $\delta z$. We use $\delta \zeta$ to refer to one of the two amplitudes: $\delta \zeta \in\{\delta x,\delta z\}$ and consider the particle's motion along the axis $\zeta \in \{x,z\}$.
The dielectric material is characterized by its complex permittivity $\epsilon_1 = \epsilon_0\varepsilon_r (1+i\tan\delta)$, with $\tan\delta$ the loss tangent and $\varepsilon_r$ the relative permittivity.
The particle, which has charge $q$, generates an electric field %\begin{equation}
$\mathbf{E}_{\zeta} =\mathbf{E}_0 + \mathrm{Re}[\mathbf{E}_{1,\zeta}e^{i\omega t}] \mathrm{,}$
%\end{equation} 
which leads to a displacement field
%\begin{align}
$\mathbf{D}_\zeta = \epsilon_1 \mathbf{E}_\zeta = \mathbf{D}_0 + \mathrm{Re}[\mathbf{D}_{1,\zeta}e^{i\omega t}] $
%\end{align} 
inside the dielectric.
Note that $\mathbf{E}_0$ and $\mathbf{D}_0$ correspond to the fields arising from the average particle position, whereas $\mathbf{E}_{1,\zeta}$ and $ \mathbf{D}_{1,\zeta}$ are due to the motion of the particle with amplitude $\delta \zeta$. 

Losses inside the dielectric material are described by the power loss of the displacement field as~\cite{Kumph_2016}
\begin{align}
\big < P_\mathrm{loss} \big >  = \int_V \frac{1}{2} \mathrm{Re}\big \{-i\omega\mathbf{E}_{1,\zeta}^* \cdot \mathbf{D}_{1,\zeta}\big \}=  \frac{\omega}{2} \epsilon_0\varepsilon_r \tan\delta \int_V |\mathbf{E}_{1,\zeta}|^2 \mathrm{.}\label{eq:loss}
\end{align}
Here, the power loss is averaged over an oscillation period of the particle, and the integral is evaluated over the volume $V$ of the dielectric.
The power loss is related to the particle motion via~\cite{Kumph_2016} \begin{equation}
\big < P_\mathrm{loss} \big > = \frac{1}{2}m \omega^2 \delta \zeta^2 \mathrm{Re}\{\gamma[\omega]\}, \label{eq:oscillation}
\end{equation}
where $\gamma$ denotes a frequency-dependent damping term.

By combining Eqs.~\ref{eq:loss} and~\ref{eq:oscillation} with the fluctuation-dissipation theorem as expressed in Ref.~\cite{Kumph_2016}, we obtain an expression for the power spectral density 
%\begin{align}
%S_\mathrm{E}(\omega) = \frac{4 k_\mathrm{B} T m \mathrm{Re}\{\gamma[\omega]\}}{q^2} \mathrm{,}
%\end{align}
\begin{align}
S_\mathrm{E}(\omega) = \frac{4 k_B T}{ \delta \zeta^2 q^2 \omega}\epsilon_0 \varepsilon_\mathrm{r} \tan \delta \int_V |\mathbf{E}_{1,\zeta}|^2 \label{eq:dissipation}
\end{align} 
of the electric-field fluctuations emerging from the dielectric at the particle position; these fluctuations are along the axis of the particle motion.
Here, $k_\mathrm{B}$ is the Boltzmann constant and $T$ the temperature of the dielectric.
Equation~\ref{eq:dissipation} thus relates energy dissipation inside the dielectric, quantified by $\mathbf{E}_{1,\zeta}$, to the electric field fluctuations originating from the dielectric.
The equation can be evaluated numerically by means of finite-element analysis (FEA) software that solves Maxwell's equations for arbitrary geometries.

%\section{\label{sec:Setup}Experimental setup}
%Before we apply the previously described formalism to predict the electric-field noise in our experimental apparatus, we first provide details on the setup. 
%To test our model, we have adapted a setup designed for quantum networking, which will be described thoroughly in Ref.~\cite{Setup}. Here, we discuss the parts of the apparatus, depicted in Fig.~\ref{fig:Fibers}, which are relevant for the simulation or the measurement of the electric-field noise.
We now turn to the experimental setup used to test our model, adapted from a setup designed for quantum networking \cite{Setup} and depicted in Fig.~\ref{fig:Fibers}.
\begin{figure}
\includegraphics[scale=0.75]{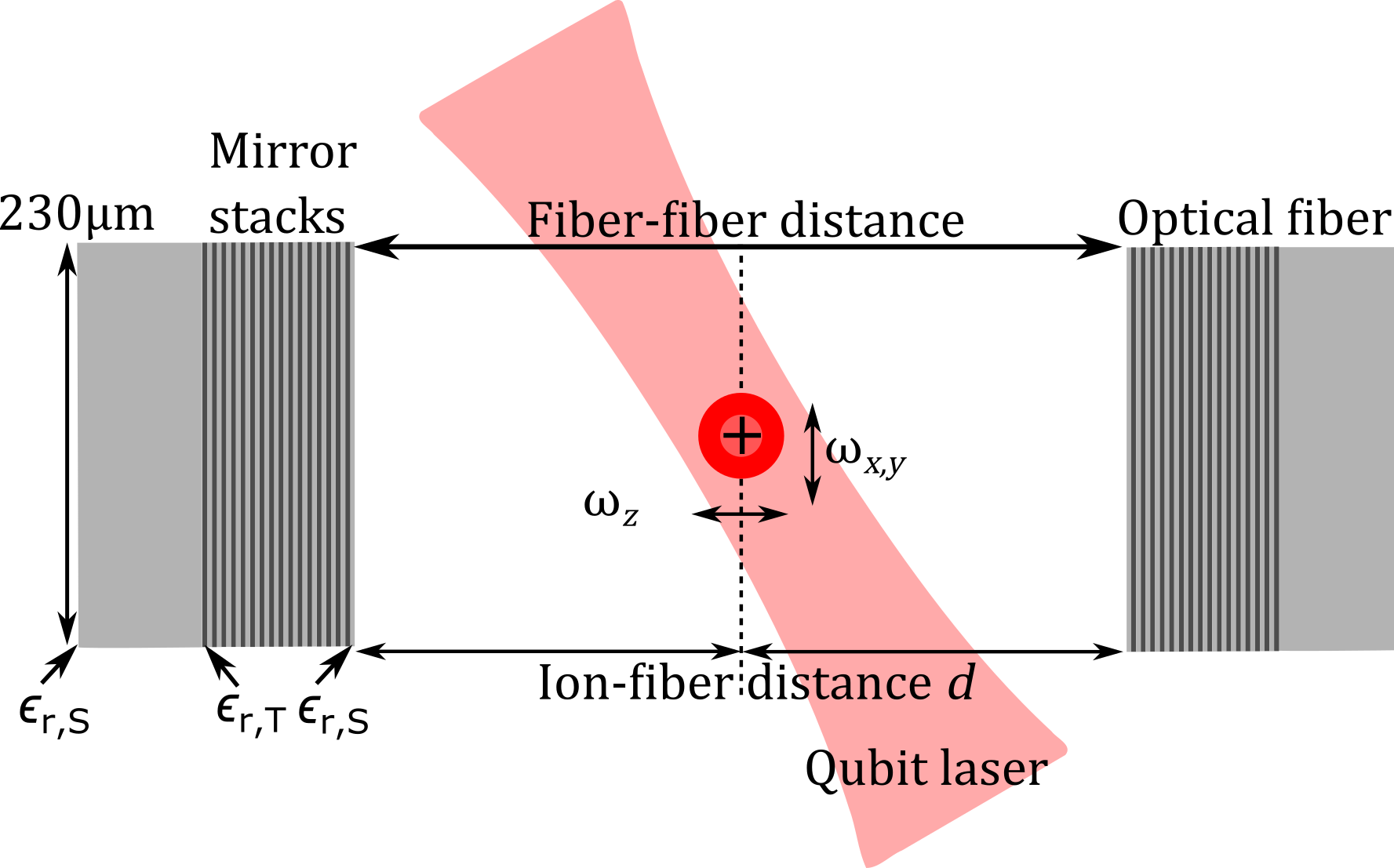}\caption{%Illustration of the experimental setup. 
	A trapped ion with axial and radial frequencies $\omega_{z}$ and $\omega_{x,y}$  is placed in the vicinity of two optical fibers. The facets of the optical fibers are laser-ablated and coated with dielectric stacks, composed of alternating layers of SiO$_2$ and Ta$_2$O$_5$, with respective permittivities $\varepsilon_\mathrm{r,S}$ and $\varepsilon_\mathrm{r,T}$. For our simulations, we neglect the mirror radii of curvature, which are $\SI{295(10)}{\micro\meter}$ for the photonic-crystal fiber and $\SI{312(5)}{\micro\meter}$ for the multi-mode fiber.}\label{fig:Fibers}
\end{figure}
A $^{40}$Ca$^{+}$ ion, trapped in a three-dimensional linear Paul trap at room temperature, is located in the center of a fiber-based optical cavity~\cite{Setup,Steinmetz06}. The trap has a radial ion-electrode distance of $\SI{250}{\micro\meter}$. %, and the heating rates of an ion in the trap was measured before the trap was integrated with the fiber cavity~\cite{Setup}. 
For the measurements presented in this paper, the axial secular frequency $\omega_{z}$ is varied from $2\pi\cdot\SI{0.5}{\mega\hertz}$ to $2\pi\cdot\SI{1.6}{\mega\hertz}$, while the radial secular frequencies are kept at $\omega_{x,y} \approx 2\pi\cdot\SI{3.3}{\mega\hertz}$.
The fiber-based optical cavity is formed by two highly reflective mirrors imprinted on the facets of a photonic-crystal fiber and a multi-mode fiber~\cite{Ott2016a}.
The mirror on the photonic-crystal fiber is composed of 41 alternating layers of ion-beam-sputtered SiO$_2$ and Ta$_2$O$_5$; the mirror on the multi-mode fiber is composed of 47 alternating layers. For both fiber mirrors, the top layer is SiO$_2$.  
The thickness of each layer corresponds to a quarter of the operating cavity wavelength of $\lambda = \SI{854}{\nano\meter}$.
%The mirrors were annealed at a temperature around $\SI{300}{\celsius}$.
The fiber-cavity length is determined from measurements of the free spectral range, which have an uncertainty of $\SI{8}{\micro\meter}$. The length can be varied from $\SI{500}{\micro\meter}$ to $\SI{1200}{\micro\meter}$ by means of piezo stages with a resolution of $\SI{1}{\nano\meter}$ for relative fiber movements. %The range of fiber separations is limited by physical constraints in the experimental setup.
We have insufficient optical access to determine the absolute position of the ion precisely with respect to each fiber. The ion-fiber distance $d$ is estimated to be half the fiber-fiber distance and therefore has a systematic uncertainty, determined from the Rayleigh length of the cavity to be $\SI{55}{\micro\meter}$~\cite{Setup}.

%\section{\label{sec:Method}Method}
%We now present a method using the previously described formalism to 
Using the formalism described above, we predict the electric-field noise originating from the dielectric in our experimental apparatus.
In our FEA simulations, each fiber is modeled as a $\SI{300}{\micro\meter}$-long cylinder with a diameter of $\SI{230}{\micro\meter}$. Each layer of the mirror coating is modeled as a cylinder with $\SI{250}{\nano\meter}$ thickness and the same diameter as the fiber. The layer thickness of $\SI{250}{\nano\meter}$ is the closest value to $\lambda/4$ that we can resolve with the FEA simulation mesh. The fiber is assumed to be at room temperature, $T= \SI{300}{\kelvin}$. For the relative permittivities of SiO$_2$ and Ta$_2$O$_5$, we use literature values of $\varepsilon_\mathrm{r,S} = 3.9$, with a loss tangent $\tan\delta_\mathrm{S} = 1.3\cdot 10^{-3}$, and $\varepsilon_\mathrm{r,T} = 22$, with $\tan\delta_\mathrm{T} = 7\cdot 10^{-3}$~\cite{Pushkar,Kim}, respectively.

%We evaluate Eq.~\ref{eq:dissipation} for the geometry depicted in Fig.~\ref{fig:Fibers} using FEA software, simulating each fiber with its mirror coating separately and afterwards summing the two noise contributions. Each fiber is modeled as a $\SI{300}{\micro\meter}$-long cylinder with a diameter of $\SI{230}{\micro\meter}$. Each layer of the mirror coating is modeled as a cylinder with $\SI{250}{\nano\meter}$ thickness and the same diameter as the fiber. The layer thickness of $\SI{250}{\nano\meter}$ is the closest value to $\lambda/4$ that we can resolve with the FEA simulation mesh. The fiber is assumed to be at room temperature, $T= \SI{300}{\kelvin}$.

%To predict the electric-field noise of our geometry, we take the following steps. 
We evaluate Eq.~\ref{eq:dissipation} for each mirror-coated fiber separately and sum the two noise contributions.
We take the calcium ion to be the point charge of Fig.~\ref{fig:Schematic}, and our task is to solve for the stationary field $\mathbf{E}_0$ inside the dielectric of Fig.~\ref{fig:Fibers}.
%In the FEA simulations, we consider a point charge at distance $d$ from the fiber's mirror coating and solve for the stationary field $\mathbf{E_0}$ inside our dielectric.
The axes for the ion's radial and axial oscillations are shown in Figs.~\ref{fig:Schematic} and~\ref{fig:Fibers}. By displacing the particle by $\delta \zeta = \SI{5}{\micro\meter}$ along a given axis and solving again for the field $\mathbf{E}_\zeta$, we determine the field $\mathbf{E}_{1,\zeta} = \mathbf{E_\zeta}-\mathbf{E}_0$ due to the oscillation of the particle along that axis. Note that the axis of the particle displacement defines the axis of the simulated electric-field noise. As long as the condition $\delta \zeta \ll d$ is fulfilled, the choice of $\SI{5}{\micro\meter}$ for $\delta \zeta$ is somewhat arbitrary since in Eq.~\ref{eq:dissipation}, the integral of $|\mathbf{E}_1|^2$ is divided by $\delta \zeta^2$, and thus $S_\mathrm{E}(\omega)$ is independent of the displacement value. However, in the FEA simulation, $\delta \zeta$ has to be small enough to avoid effects from the borders of the simulated geometry yet larger than the meshing resolution of our simulated geometry. We found the simulated electric-field noise to be independent of $\delta \zeta$ for displacements between $\SI{5}{\micro\meter}$ and $\SI{25}{\micro\meter}$ for all ion-dielectric separations considered in this work.

In the following experiments, we investigate whether the rate at which the ion's temperature increases is supported by our predictions of the electric-field noise emerging from the mirror-coated fibers. We measure the motional heating rate using the decay of carrier Rabi oscillations due to the Debye-Waller coupling~\cite{Roos,10.5555/2011477.2011478,HAFFNER2008155,PhysRevA.100.063405}. 
After Doppler cooling and preparation of the ion in the $|4^2S_{1/2}, m_j = -1/2 \rangle$ state, we wait for a time $t$ and then excite the $|4^2S_{1/2}, m_j = -1/2 \rangle$ to $|3^2D_{5/2}, m_j=-1/2\rangle$ $^{40}$Ca$^{+}$ transition using a narrow-linewidth laser with wavelength $\lambda = \SI{729}{\nano\meter}$. The laser beam overlaps with the ion's radial and axial modes of motion and therefore probes the ion's temperature along all three axes simultaneously~\cite{Wineland1998}.
From a fit of Rabi oscillations in which each data point is weighted according to the quantum projection noise, we extract the parameter\begin{equation}
\beta = \sum_i \bar{n}_i \eta_i^2 \mathrm{,}
\end{equation} where $\bar{n}_i$ is the mean phonon number along the motional axis $i\in\{x,y,z\}$ and $\eta_i = \frac{2\pi}{\lambda}\sqrt{\frac{\hbar}{2m\omega_i}}\cos(\phi_i)$ denotes the effective Lamb-Dicke factor of each mode, with $\phi_i$ the angle between the mode axis and the wave vector~\cite{Roos}.
We determine $\beta$ for several values of $t$ and deduce $\dot{\beta}$ from a linear fit. 
We identify $\dot{\beta}$ as a weighted sum of the heating rates $\dot{\bar{n}}_i$ along the three motional axes: $\dot{\beta} = \sum_i \dot{\bar{n}}_i \eta_i^2 \mathrm{.}$
The uncertainty of $\dot{\beta}$ corresponds to the standard deviation of this fit.

In order to compare experimental observations to the predictions of our model, we calculate the heating rate along each motional axis that results from the simulated electric-field noise~\cite{RevModPhys.87.1419}: \begin{equation}
\dot{\bar{n}}_i = \frac{q^2}{4 m \hbar \omega}S_\mathrm{E}(\omega_i) \mathrm{.}  \label{eq:Heating_rate}
\end{equation} Thus, we predict the heating rates for all three axes separately, allowing us to determine $\dot{\beta}$.

For our experimental geometry, the axial contribution to the heating rate dominates by more than one order of magnitude~\footnote{\label{note1}See Supplemental Material at [URL  will be inserted by publisher] for more details on fitting of the distance scaling, testing the simulation method with an infinite plane, and determining the loss tangent from the frequency scaling, as well as for a table of axial trap frequencies corresponding to the data of Fig.~\ref{fig:Distance}.}. Thus, the heating rate along the $z$ axis in the system is approximately equal to (though smaller than) $\dot{\beta}/\eta_z^2$.
For easier comparison with observations of other experimental platforms, which typically consider heating of just one mode at a time, we define 
%using the expression
%\begin{equation}
$\dot{\bar{n}} = \dot{\beta}/\eta_z^2 $
as a projection onto the $z$ axis .
%\label{eq:Heating_rate_scale}
%\end{equation} 
% Eq.~\ref{eq:Heating_rate_scale}.

We first study the frequency response of the heating rate. For these measurements, we vary $\omega_z$ for two ion-fiber separations of $\SI{450}{\micro\meter}$ and $\SI{600}{\micro\meter}$. 
\begin{figure}
\includegraphics[scale=0.55]{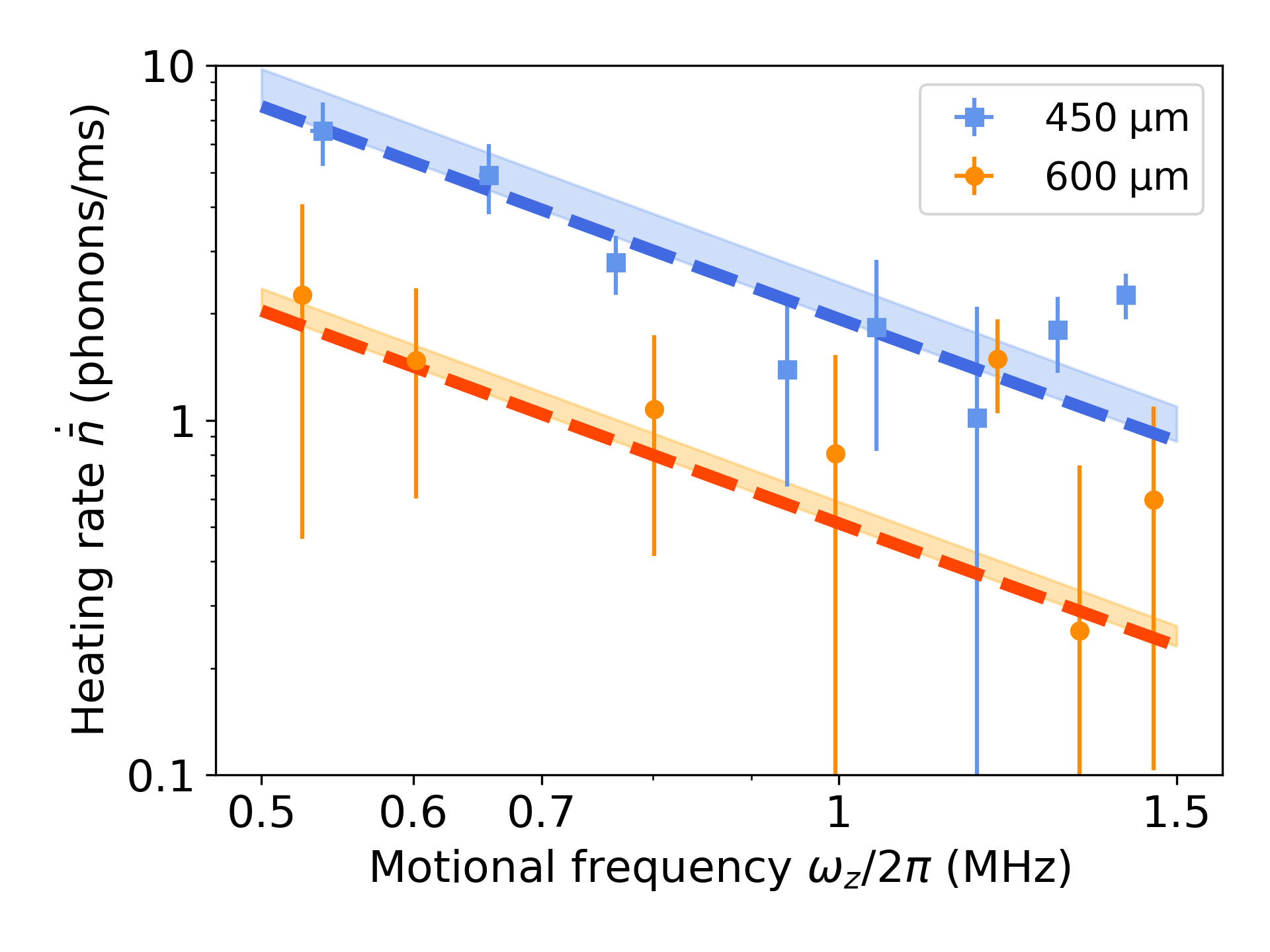}
\caption{The heating rate as a function of axial frequency for ion-dielectric separations of $\SI{450}{\micro\meter}$ (squares) and $\SI{600}{\micro\meter}$ (circles). The dashed lines represent the predictions based on the literature values, and light-orange and light-blue areas show systematic uncertainties of the ion-fiber distance.}
\label{fig:Frequency}
\end{figure}
In Fig.~\ref{fig:Frequency}, we plot measured data together with predictions from simulations based on literature values for the dielectric material properties, with no free parameters. Over a frequency range from $2\pi\cdot\SI{0.5}{\mega\hertz}$ to $2\pi\cdot\SI{1.5}{\mega\hertz}$, the heating rate varies by an order of magnitude for both separations, which is captured accurately by our FEA-based noise estimation. We calculate a reduced chi-square value of $\chi_\mathrm{\nu}^2 = 1.39$ for experiment and predictions, for both data sets combined. Given this agreement, an interesting application is that our method can be employed to determine the loss tangent of unknown dielectric materials~\cite{Note1}.

Second, we measure the distance scaling of the heating rate, varying $d$ while keeping the ion position fixed within $\SI{25}{\micro\meter}$ using a CCD camera image as reference.
For each distance, we measure the axial trap frequency $\omega_z/2\pi$, which ranges from $\SI{1.300(8)}{\mega\hertz}$ to $\SI{1.696(3)}{\mega\hertz}$. This measurement is necessary as the trap potential changes with $d$ due to surface charges on the dielectric. 
In Fig~\ref{fig:Distance}, we show the experimental data and the predicted heating rates from our model without free parameters. The predicted rates are plotted as discrete values evaluated for each measured ion-fiber separation and trap frequency. Additionally, we fit the eight simulated electric-field-noise values with a power-law function $S_\mathrm{E} = A\cdot d^{-\alpha}$~\cite{Note1}.  The radial and axial scaling parameters are determined to be $\alpha_\mathrm{rad} = 4.413(6)$ and $\alpha_\mathrm{ax} = 4.016(6)$. The divergence of these parameters from the $1/d^3$ scaling of an infinite dielectric plane underscores the necessity of our method for finite geometries~\cite{Henkel1999}. To visualize the scaling for a single trap frequency, we plot the heating rate based on the fitted electric-field noise as a continuous function for $\omega_z = 2\pi \cdot \SI{1.6}{\mega\hertz}$.

%We determine the electric-field noise for discrete ion-fiber distances with the FEA simulations and subsequently interpolate between these distances with a power-law function $S_\mathrm{E} = A\cdot d^{-\alpha}$. The discrete values of the electric-field noise and the corresponding power-law functions are depicted in Fig.~\ref{fig:Simulation2}. From the power-law functions, we determine the scaling parameters of electric-field noise along the radial and axial mode of motion to be $\alpha_\mathrm{rad} = 4.413(6)$ and $\alpha_\mathrm{ax} = 4.016(6)$.  Using this power-law functions, we predict the heating rates as a continuous function of the ion-fiber separation.

 %as the trap potential changes due to surface charges on the dielectric, subsequently adjusting the voltage on the trap electrodes to achieve $\omega_\mathrm{z} \approx 2\pi\cdot\SI{1.6}{\mega\hertz}$.
%The axial trap frequency for each ion position is listed in Tab.~\ref{Tab:appendix} in the appendix. 
\begin{figure}
\includegraphics[scale=0.5]{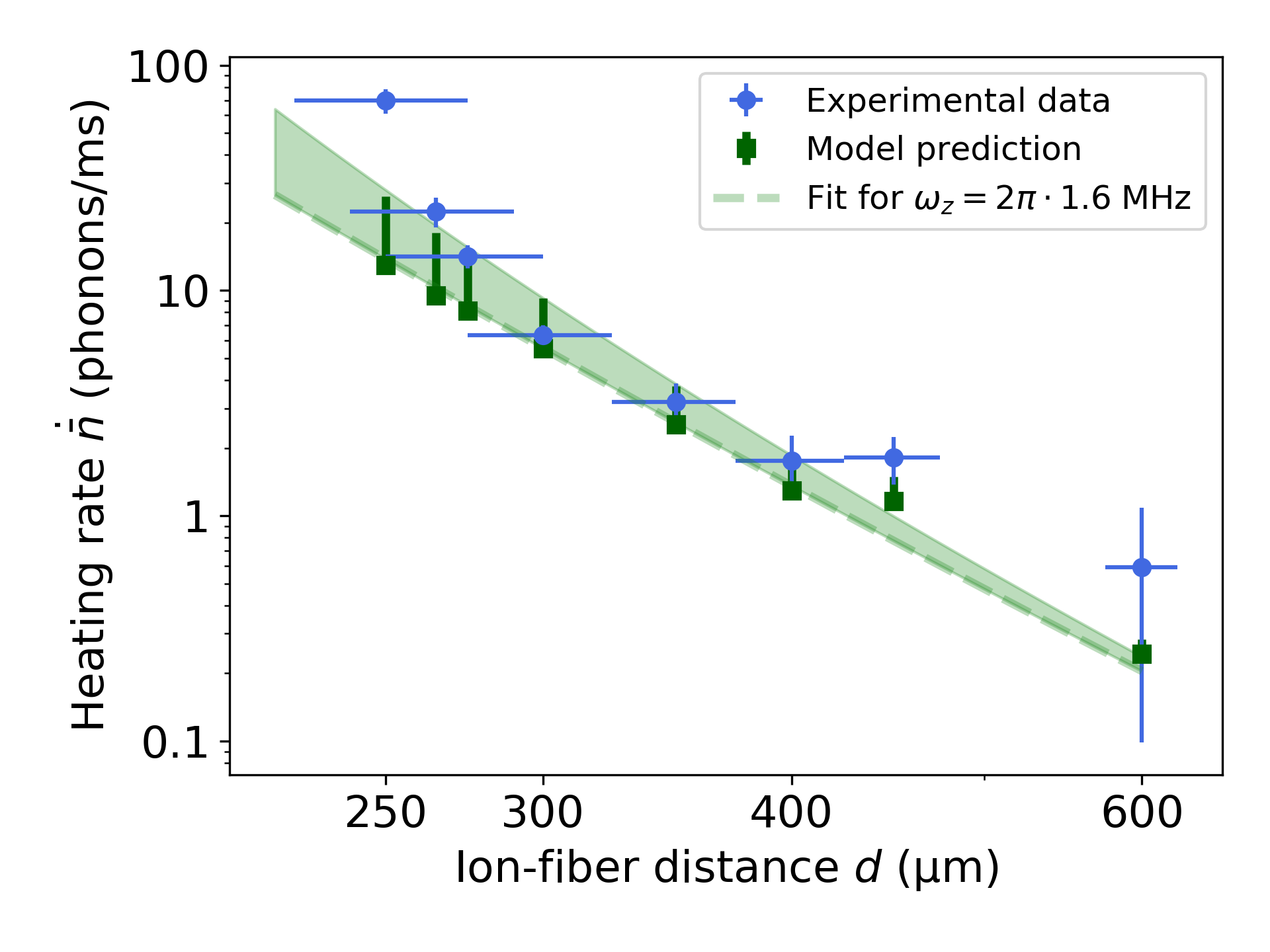}
\caption{The heating rate as a function of ion-fiber distance. The green points represents the prediction with errorbars corresponding to the systematic uncertainty of the ion-fiber distance. 
The dashed line depicts a fit for a trap frequency of $\omega_z = 2\pi \cdot \SI{1.6}{\mega\hertz}$, and the respective systematic uncertainty is illustrated in light green.}\label{fig:Distance}
\end{figure}

For the full set of data shown in Fig.~\ref{fig:Distance}, from $d=\SI{250}{\micro\meter}$ to $d=\SI{600}{\micro\meter}$, we calculate $\chi_\mathrm{\nu}^2=2.57$ for the measurements and the model.  Excluding the point corresponding to the shortest distance of $d = \SI{250}{\micro\meter}$ results in a value of $\chi_\mathrm{\nu}^2 = 0.86$, demonstrating strong agreement.
At this short distance, the axial trapping potential is significantly influenced by inhomogeneously distributed surface charges, which, however, are not considered in the simulations~\cite{Ong2020probing}. Potentially, more information could be extracted by measuring a frequency scaling similar to that of Fig.~\ref{fig:Frequency}.
However, below $d = \SI{300}{\micro\meter}$, the ion is observed to be unstable for $\omega_\mathrm{z} \leq 2\pi\cdot\SI{1.5}{\mega\hertz}$, impeding such a measurement. 
%At $\SI{250(25)}{\micro\meter}$, a discrepancy emerges.
%At these short distances, the axial trapping potential is significantly influenced by surface charges, which, however, are not considered in the simulations~\cite{Ong2020probing}.
%We suspect that an inhomogeneous distribution of these charges induces anharmonicities in the trapping potential.
%Potentially, more information on the noise source %at these distances 
%could be extracted by measuring a frequency scaling similar to that of Fig.~\ref{fig:Frequency}.
%However, below $d = \SI{300}{\micro\meter}$, we observe that the ion is unstable for $\omega_\mathrm{z} \leq 2\pi\cdot\SI{1.5}{\mega\hertz}$ and lost, impeding a measurement of frequency scaling.
%However, below $d = \SI{300}{\micro\meter}$, we observe that the ion is unstable for $\omega_\mathrm{z} \leq 2\pi\cdot\SI{1.5}{\mega\hertz}$ and no longer crystallizes, impeding a measurement of frequency scaling.

%\section{\label{sec:Context}Context}

%In the following section, we place our results in context of previous studies and discuss the implications of this work on the design of novel ion trap apparatuses.

In summary, we have presented a new FEA-based method to predict electric field noise from dielectrics with arbitrary geometry. We have taken advantage of an ion-trap experiment with a movable dielectric element in order to test this method. The measured heating rates are attributed to the influence of the bulk dielectric as they are up to three orders of magnitude larger than reference values measured before the trap was integrated with the cavity: 
$13(3)\;\mathrm{phonons/s}$ for the axial mode of motion at a frequency of $2\pi\cdot \SI{1.5}{\mega\hertz}$ and $\{26(6),32(8) \}$ phonons/s for the radial modes at frequencies of $2\pi\cdot\{ 3.2,3.4\}\;\mathrm{MHz}$~\cite{Setup}. While other sources of technical noise may be present, we do not expect the heating rates due to these sources to scale with the distance to the fiber. The magnitude of the electric-field noise predicted by our model without free parameters agrees with the measured data for a range of motional frequencies and ion-dielectric separations. The analysis for the ion-dielectric separation of $\SI{250}{\micro\meter}$ and below will require further study. 
%Our observations emphasize the contribution of dielectric materials to anomalous heating in ion traps. 

Here we have focused on the effects of a bulk dielectric, but we emphasize that the contributions of dielectric materials to heating in ion traps may be a much more common phenomenon. As an example, we predict the influence of an SiO$_2$ cylinder with radius of $\SI{10}{\micro\meter}$ and a thickness of only one nanometer. Such a structure could appear as surface contamination during the fabrication of an ion trap. Assuming an ion-dielectric separation of $\SI{50}{\micro\meter}$ and trap frequencies of $\omega = 2\pi\cdot \SI{1}{\mega\hertz}$ for all three modes of a $^{40}$Ca$^+$ ion, we find heating rates ranging from $127\;\mathrm{phonons/s}$ to $170\;\mathrm{phonons/s}$ at room temperature. We thus see that nanoscale patches of surface oxides may lead to significant heating when compared to the target rates for scalable ion-trap quantum computing and communication platforms~\cite{PhysRevX.7.041061}. 

Our prediction method can be used to systematically optimize the design and fabrication of future trapped-ion setups, as the electrode geometry and materials --- including contaminant or oxide layers --- can be adjusted in the FEA simulations to suppress heating effects due to dielectrics.
Finally, we note that these results are also relevant for devices such as nitrogen-vacancy centers~\cite{PhysRevLett.118.197201,PhysRevApplied.10.064056} and donor spins in silicon~\cite{Muhonen2014} that are disturbed by environmental noise due to nearby surfaces.

%\section{\label{sec:Conclusion}Conclusion}

%In summary, we have presented a method that predicts electric-field noise due to bulk dielectric materials. We have experimentally probed this kind of surface noise  by means of a trapped ion. Furthermore, by comparing the experimental measurements to our method's predictions, we have extracted the loss tangent of our dielectric material. Our method can be used during the design process of ion traps with integrated dielectric elements to minimize the motional heating due to the dielectric.

%TC:ignore
All data presented and discussed in this manuscript are available at https://doi.org/10.5281/zenodo.4600786.

We thank Carsten Henkel for his feedback on the infinite plane model and for his suggestion to use the fluctuation-dissipation theorem to simulate heating rates. We thank Brenda Rubenstein and Yuan Liu for  interesting and fruitful discussions.
This work was supported by the European Union's Horizon 2020 research and innovation program under Grant Agreement No. 820445 (Quantum Internet Alliance), by the U.S. Army Research Laboratory under Cooperative Agreement No. W911NF-15-2-0060, and by the Austrian Science Fund (FWF) under Project No. F 7109. P.H. acknowledges support from the European Union's Horizon 2020 research and innovation programme under Grant Agreement No. 801285 (PIEDMONS). P.S. acknowledges support from the Austrian Research Promotion Agency (FFG) contract 872766.

This material is based upon work supported by the Department of Defense under Air Force Contract No. FA8702-15-D-0001. Any opinions, findings, conclusions, or recommendations expressed in this material are those of the authors and do not necessarily reflect the views of the Department of Defense.
\bibliography{References}
\clearpage

\appendix
\title{Supplemental Material:\\ Heating of a trapped ion induced by dielectric materials}
\maketitle
\section{Testing the simulation method with an infinite plane}
In this section, we %verify that our method accurately predicts electric-field noise from dielectric materials. We 
predict the electric-field noise due to a thin cylinder of dielectric and compare the results to an infinite-plane solution based on the established fluctation electrodynamics (FE) formalism~\cite{PhysRevA.30.1185,Henkel1999}. Here, the term infinite plane refers to a dielectric half-space with a planar dielectric-vacuum interface.

Following the steps outlined %for our method 
in the main text, we determine the spectral density of the electric-field noise as a function of the ion-dielectric separation for a plane of SiO$_2$. The plane that we simulate via finite-element analysis (FEA) needs to have finite boundaries, so we model an infinite plane as a cylinder with diameter of $\SI{6}{\milli\meter}$ and a thickness of $\SI{250}{\micro\meter}$.
\begin{figure}
\includegraphics[scale=0.5]{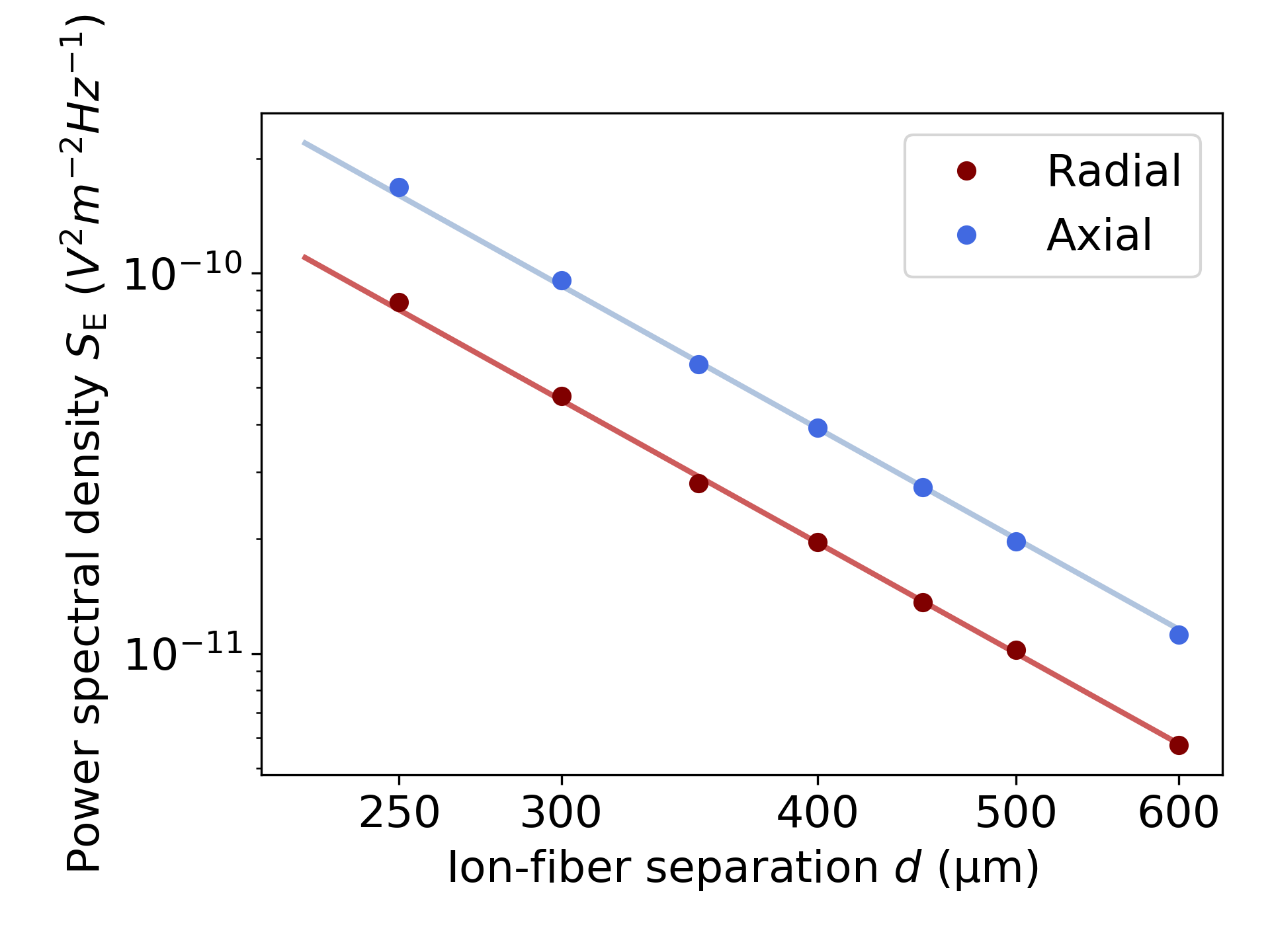}
\caption{The electric-field noise $S_E$ as a function of the ion-dielectric separation along both axial and radial ion-trap axes.  The points show FEA-based simulations for a thin cylinder approximating an infinite plane. The solid lines show the numerical evaluation of Eq.~13 in Ref.~\cite{Henkel1999} for an infinite plane. We choose both the axial and radial motional frequencies to be $2\pi\cdot\SI{1}{\mega\hertz}$. %The scaling parameters $\alpha$ for each curve are listed in Tab.~\ref{tab:scaling}.
}\label{fig:Simulation}
\end{figure}
In Fig.~\ref{fig:Simulation}, we plot the simulated electric-field noise along the axial and radial ion-trap axes indicated in Figs.~1 and~2 of the main text. For comparison, we plot the numerical evaluation of Eq.~13 in Ref.~\cite{Henkel1999}, which is a solution for the electric-field noise based on an infinite dielectric plane. 

Based on the data of Fig.~\ref{fig:Simulation}, one sees that the two methods are in agreement in this case.  We proceed in the main text to extend our approach to dielectric materials of arbitrary geometry, which the FE formalism is unable to address.
%In contrast to the simulations, the analytic solution does not take geometrical restrictions into account and considers the first two layers of the mirror coating as infinite planes, for which the Maxwell equations can still be evaluated analytically~\cite{PhysRevA.30.1185,Henkel1999}.

\section{Interpolation of the distance scaling} \label{sec:SupDist}
\begin{figure}
\includegraphics[scale=0.5]{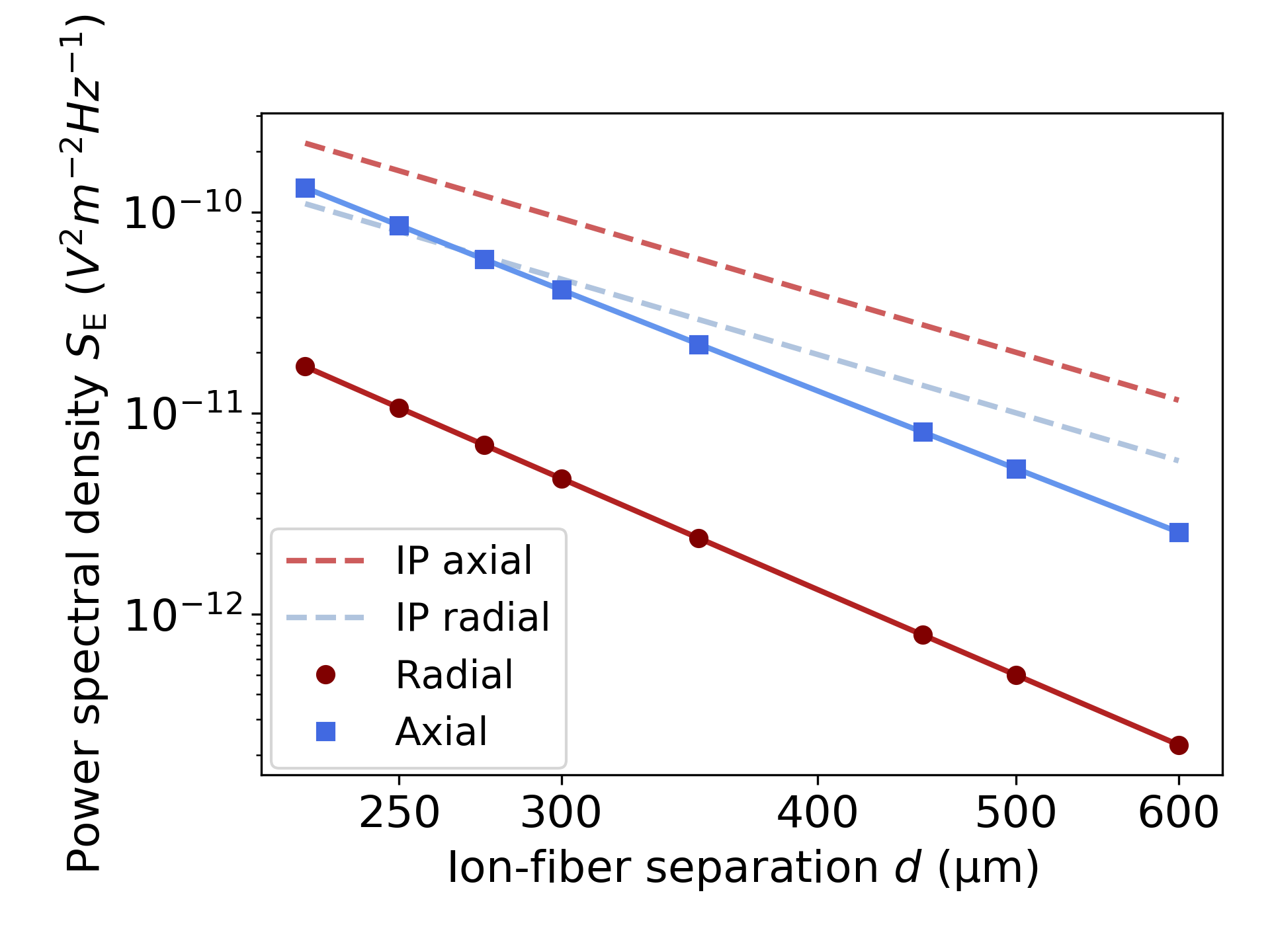}
\caption{Simulation of the power spectral density $S_E$ at the ion location as a function of the ion-fiber separation $d$, based on the experimental configuration indicated in Fig.~2 of the main text. We choose the axial trap frequency to be $\omega_{z} = 2\pi\cdot\SI{1}{\mega\hertz}$ and the radial trap frequency to be $\omega_{x,y} = 2\pi \cdot \SI{3.3}{\mega\hertz}$.  The discrete points are the simulation results, and the solid lines are fits to a power law $S_\mathrm{E} = A\cdot d^{-\alpha}$. For comparison, we plot the results of a model in which the fiber is replaced by an infinite plane (IP). %The scaling parameters $\alpha$ for each curve are listed in Tab.~\ref{tab:scaling}.
}\label{fig:Simulation2}
\end{figure}

In the FEA simulation, we determine the electric-field noise for discrete values of the ion-fiber separation. To determine the electric-field noise for distances between these values, we fit a power-law function $S_\mathrm{E} = A\cdot d^{-\alpha}$ to the results. For the optical fibers described in the main text, both the simulated electric-field noise and the fitted power-law functions along the axial and radial modes of motion are depicted in Fig.~\ref{fig:Simulation2}. The trap frequencies are chosen to be $\omega_{z}=2\pi\cdot\SI{1}{\mega\hertz}$ and $\omega_{x,y} = 2\pi \cdot \SI{3.3}{\mega\hertz}$. 
From the fits, we extract the radial and axial scaling parameters to be $\alpha_\mathrm{rad} = 4.413(6)$ and $\alpha_\mathrm{ax} = 4.016(6)$. In Fig.~\ref{fig:Simulation2}, we also see that the axial contribution to the power spectral density is around one order of magnitude larger than the radial contribution.

For comparison, we plot the electric-field noise for an infinite plane of SiO$_2$ described by Eq.~13 of Ref.~\cite{Henkel1999}, which follows a $1/d^3$ scaling.
The power spectral densities corresponding to the two models differ in magnitude and shape, quantified by the scaling parameters of the curves.
Although the physical origin of the noise is identical for both simulations, we observe a significant difference in the scaling parameters, consistent with the conclusion in Ref.~\cite{PhysRevA.84.053425} that the distance scaling of a noise model is strongly dependent on geometrical restrictions.
Because of this discrepancy, in the main text, we compare measured data only with simulations based on our experimental geometry.

\section{Determining the loss tangent from the frequency scaling}

\begin{figure}
	\includegraphics[scale=0.55]{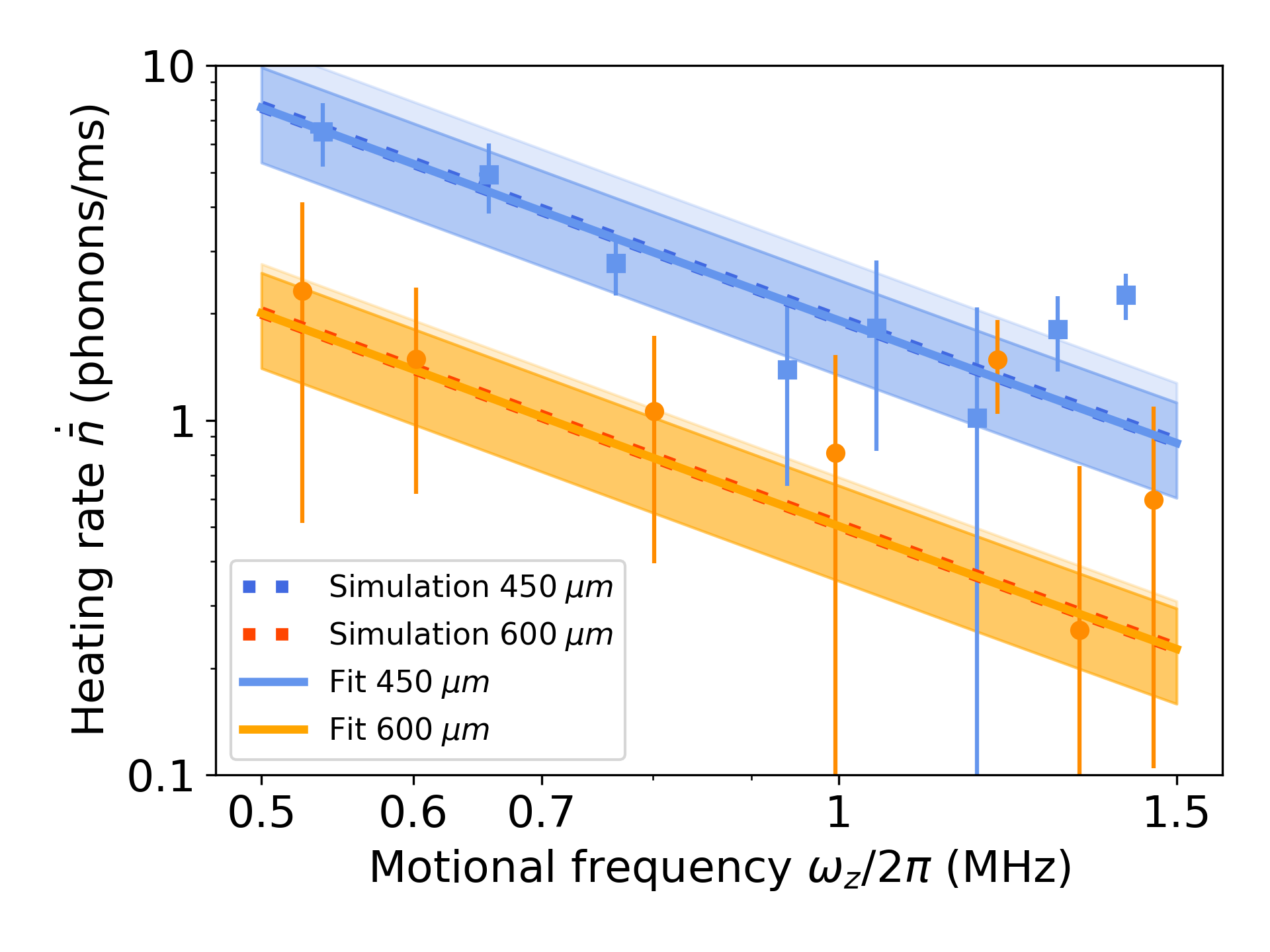}
	\caption{The heating rate as a function of axial frequency for ion-dielectric separations of $\SI{450}{\micro\meter}$ (squares) and $\SI{600}{\micro\meter}$ (circles). The dashed lines represent the predictions based on the literature values, and the solid lines illustrate fits of Eq.~3 in the main text, with the loss tangent as a common free parameter. The  dark orange and dark blue areas illustrate the uncertainty of the fitted loss tangent. Light orange and light blue areas show systematic uncertainties of the ion-fiber distance.}
	\label{fig:SupFrequency}
\end{figure}

The loss tangent is a material property that varies from sample to sample~\cite{Kim}, and its exact value depends on the details of the manufacturing process~\cite{Wang}, the sample temperature~\cite{MicrowaveElectronics2} and the electric-field frequency~\cite{MicrowaveElectronics1}.  In particular, for Ta$_2$O$_5$, the loss tangent $\tan\delta_\mathrm{T} = 7\cdot 10^{-3}$ from Ref.~\cite{Kim} disagrees with the value of $1.5\cdot 10^{-2}$ reported in Ref.~\cite{Pushkar}; error bars are not provided for either value. Note that both literature values have been measured at room temperature for frequencies around $\SI{1}{\mega\hertz}$, matching the regimes that we experimentally observe.

Thus, as another means of comparison, we determine from our measurements and the FEA simulations the loss tangent of Ta$_2$O$_5$.  We use the literature values for the complex permittivity of SiO$_2$ and for the relative permittivity $\epsilon_\mathrm{r,T}$ of Ta$_2$O$_5$, thus leaving the loss tangent of Ta$_2$O$_5$ as free parameter. For each motional axis, we determine Eq.~3 of the main text as a function of $\omega$ and $\tan\delta_\mathrm{T}$, using our simulated field $\mathbf{E_{1,\zeta}}$. Calculating the heating rates from Eq.~5 and weighting the rates according to the beam geometry results in the parameter $\dot{\beta}(\omega,\tan\delta_\mathrm{T})$. To determine $\tan\delta_\mathrm{T}$, we  fit both data sets in Fig.~3 of the main text simultaneously to this expression. In Fig~\ref{fig:SupFrequency}, we reproduce the data shown in Fig.~3 and additionally indicate the fit of the loss tangent as solid lines. The loss tangent of Ta$_2$O$_5$,  $\tan\delta_\mathrm{T,fit} = 6.9(2.1)\cdot 10^{-3}$, determined from the fit, agrees within one standard deviation with the literature value in Ref.~\cite{Kim}. %We calculate the reduced chi-square of $\chi_\mathrm{fit}^2 = 1.87$ for the predictions with fitted value, which indicates that the model using the literature value matches equally well to the experiment as the model with fitted parameter.
The uncertainty of the extracted loss tangent is depicted as dark blue and dark orange areas in Fig.~\ref{fig:SupFrequency}. The systematic uncertainty of the ion-fiber distance is indicated as light blue- and orange-shaded areas. 

Due to the similar contributions of Ta$_2$O$_5$ and SiO$_2$ to the heating rate, we cannot fit for the loss tangents of both materials separately. Consequently, the fit to determine $\tan\delta_\mathrm{T,fit}$ compensates for any discrepancy between the literature and experimental values of SiO$_2$. Thus, the loss tangent of Ta$_2$O$_5$ determined from the fit does not necessarily correspond to its true value but rather helps us to establish that the data are consistent with loss tangents close to literature values. Nevertheless, this approach can be applied in experimental setups with a single dielectric material to precisely determine the loss tangent.
%%\begin{figure}
%\includegraphics[scale=0.5]{figures/figure4/figure4.png}
%\caption{The heating rate as a function of ion-fiber distance.
%The solid line is the prediction based on the fitted  Ta$_2$O$_5$ loss tangent. The uncertainty due to the loss tangent is shown in dark blue and the systematic uncertainty of the ion-fiber distance in light blue.}\label{fig:SupDistance}
%\end{figure}
%\newpage
\section{Table of axial trap frequencies in Fig.~4}
\begin{table}[h]
\begin{tabular}{|c|c|}
\hline 
Distance $d$ ($\mu \mathrm{m}$) & Frequency $\omega_z/2\pi$ (MHz) \\ 
\hline 
250(25) & 1.669(5) \\
\hline 
265(25) & 1.668(5) \\  
\hline 
275(25) & 1.636(5) \\ 
\hline 
300(25) & 1.622(5) \\ 
\hline 
350(25) & 1.660(5) \\ 
\hline 
400(25) & 1.644(5) \\ 
\hline 
450(25) & 1.300(8) \\ 
\hline 
500(25) & 1.696(5) \\ 
\hline 
600(25) & 1.459(3) \\ 
\hline 
\end{tabular}
\caption{Distances and axial trap frequencies depicted in Fig.~4 of the main text.}
\label{Tab:appendix}
\end{table}

\bibliography{References}

\end{document}